\documentstyle[epsf, prl, twocolumn, aps, floats]{revtex}
\begin{document}
\twocolumn

\title{Semiclassical evidence for the BGS-conjecture}

\author{Stefan Heusler}
\address{heusler@theo-phys.uni-essen.de,
Fachbereich Physik, Universit\"at Essen,
45\,117 Essen, Germany }
\date{\today}
\maketitle

\begin{abstract}
€
Recently, M. Sieber and K. Richter achieved a breakthrough towards a proof of the BGS-conjecture by calculating a first semiclassical correction to the diagonal approximation of the orthogonal form factor for geodesic flow on a Riemann 
surface of constant negative curvature. 

In this note, we try to generalize the arguments. However, the solution proposed is not yet correct, because also other geometries must be taken into account.

\end{abstract}€€

\vspace{.4cm}


\vspace{.7cm}

\noindent {\it Introduction:}
Experimental and numerical work for many chaotic quantum systems has shown their spectral fluctuations to be  faithful to random-matrix theory \cite{ref0,stoeck}. However, up to now, the equivalence of the spectral properties of random-matrix theory with the properties of classically chaotic quantum systems has only been a conjecture \cite{refA}. Following pioneering work of Sieber and Richter \cite{ref1}, we here give analytic evidence by a semiclassical treatment of geodesic flow on a surface of constant negative curvature with genus $g \geq 2$. The origin of universality of spectral fluctuations for chaotic quantum billiards and related systems with two-dimensional configuration space seems to lie in generic properties of self-intersections of periodic orbits in configuration space. Even though derived for a special system, our results could pave the way towards understanding universality: One would have to reveal the statistical properties of self-intersections as applicable to other dynamics.

The so-called form factor, the Fourier transfom of the two-point correlator of the level density, is prediced by random-matrix theory for dynamics with time reversal invariance as \cite{ref0}

\[
K_{\rm orth} (\tau)  =
 \left\{
\begin{array}{r@{\quad:\quad}l}
2 | \tau | - | \tau | \ln [1 + 2 | \tau | ] & {\rm for} \ | \tau | < 1
\\
2 - | \tau | \ln \frac{ 2 | \tau | + 1}{ 2 | \tau| - 1} & {\rm for} \ | \tau | > 1  \ \  ,
\end{array} \right. \]
where $\tau$ is the time in units of the Heisenberg time. In the interval $0 \le \tau \le 1/2$, one has the expansion $K(\tau) = 2 \tau + \tau \sum\limits_{k=1}^{\infty} \frac{(-1)^k}{k} 2^k \tau^k $. 

Let $A_{\gamma} \equiv T_{\gamma} / \sqrt{{\rm Tr} {\rm M}_{\gamma} - 2}$ be the semiclassical amplitude of the classical periodic orbit $\gamma$ with period $T_{\gamma}$, action $S(\gamma)$, and monodromy matrix $M_{\gamma}$. In the semiclassical framework, the form factor is then given by the following double sum over periodic orbits \cite{ref0}

\begin{eqnarray}
\label{forfac}
K_{\rm orth} (\tau)  = \   \ \ \ \ \ \ \ \ \ \ \ \ \   \ \ \ \ \ \ \ \ \ \ \ \ \   
\nonumber\\
\lim_{\hbar \rightarrow 0} 
\frac{1}{T_H} \sum\limits_{\gamma, \gamma'} 
A_{\gamma} A^{*}_{\gamma'}
{\rm e}^{ {i (S_{\gamma} - S_{\gamma'})/\hbar} } \delta(T - \frac{T_{\gamma} + T_{\gamma'}}{2})  \  \ .
\end{eqnarray}
Here, $T_H$ denotes the Heisenberg time. According to Berry's 1985 insight \cite{refB}, one obtains the leading-order term $ 2 \tau $ of random-matrix theory by retaining only the diagonal terms, $\gamma = \gamma'$, as well as the contructive interference of pairs of mutually time reversed orbits. We propose to establish the full $\tau$-expansion, of which the second order term $- \ 2 \tau^2$ was found in the pioneering work of Sieber and Richter. They recognized that term as due to certain pairs of periodic orbits which are (in a certain sense exponentially) close neighbours in configuration space; one orbit in each pair has an additional self-intersection (with small angle $\epsilon$) relative to the other. A standard argument of Riemannian geometry \cite{abresch} reveals the non-intersecting satellite orbit as unique. In summing over the said pairs in equation (\ref{forfac}), we shall be led to sums over intersections.

Due to the mutual closeness of the orbits in each pair, their action difference $S(\gamma) - S(\gamma')$ is not large compared to $\hbar$; it is dominated by the behavior in the immediate neighbourhood of the intersection. In contrast to other "uncorrelated" pairs of orbits, the ones under consideration will therefore interfere constructively in the form factor. 
The pairs of orbits contributing the $\tau^2$-term in $K(\tau)$ involve one intersection. We shall show the $\tau^{k+1}$ term to arise from families of orbit pairs involving $k$ intersections (see Fig. 1).

\begin{figure}[t]
\begin{center}
\leavevmode
\epsfxsize=0.45\textwidth
\epsffile{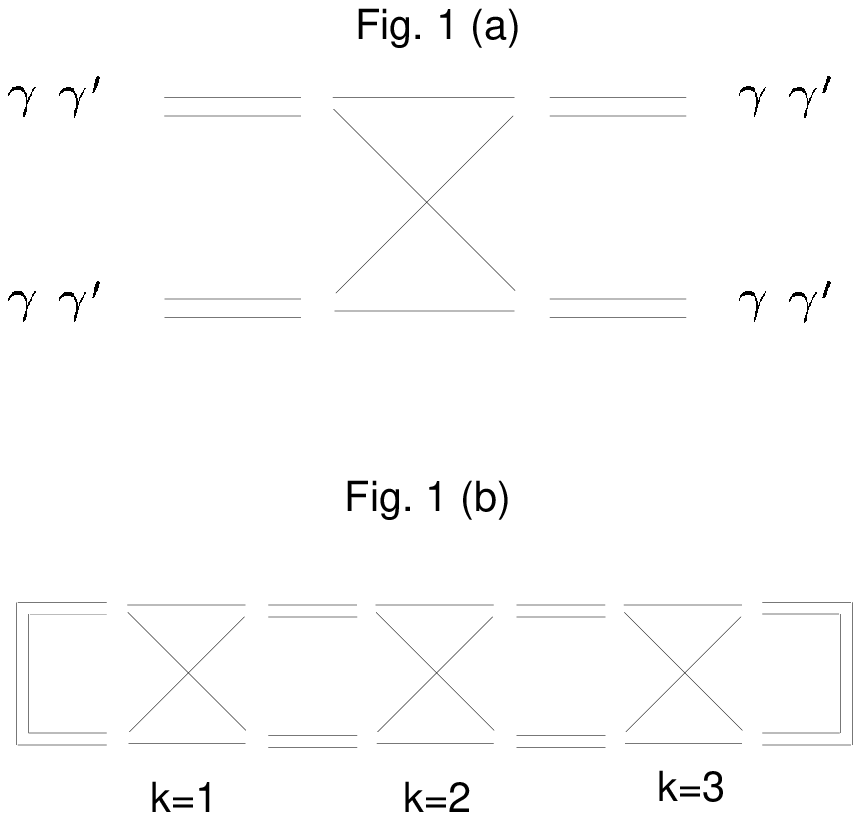}
\end{center}
\caption{
(a) The "intersection box": When one orbit in a pair self-intersects, the companion avoids the crossing. There are two possibilities: Each arriving orbit may choose to self-intersect. 
(b) A $k$-family comprises $2^k$ pairs of orbits with periods near $T$ which have in the mean $\propto T^2$ intersections. The orbits in a pair differ only in the neighbourhood of precisely $k$ points (the intersection boxes).}
\label{abbdelta}
\end{figure}

\begin{figure}[t]
\begin{center}
\leavevmode
\epsfxsize=0.45\textwidth
\epsffile{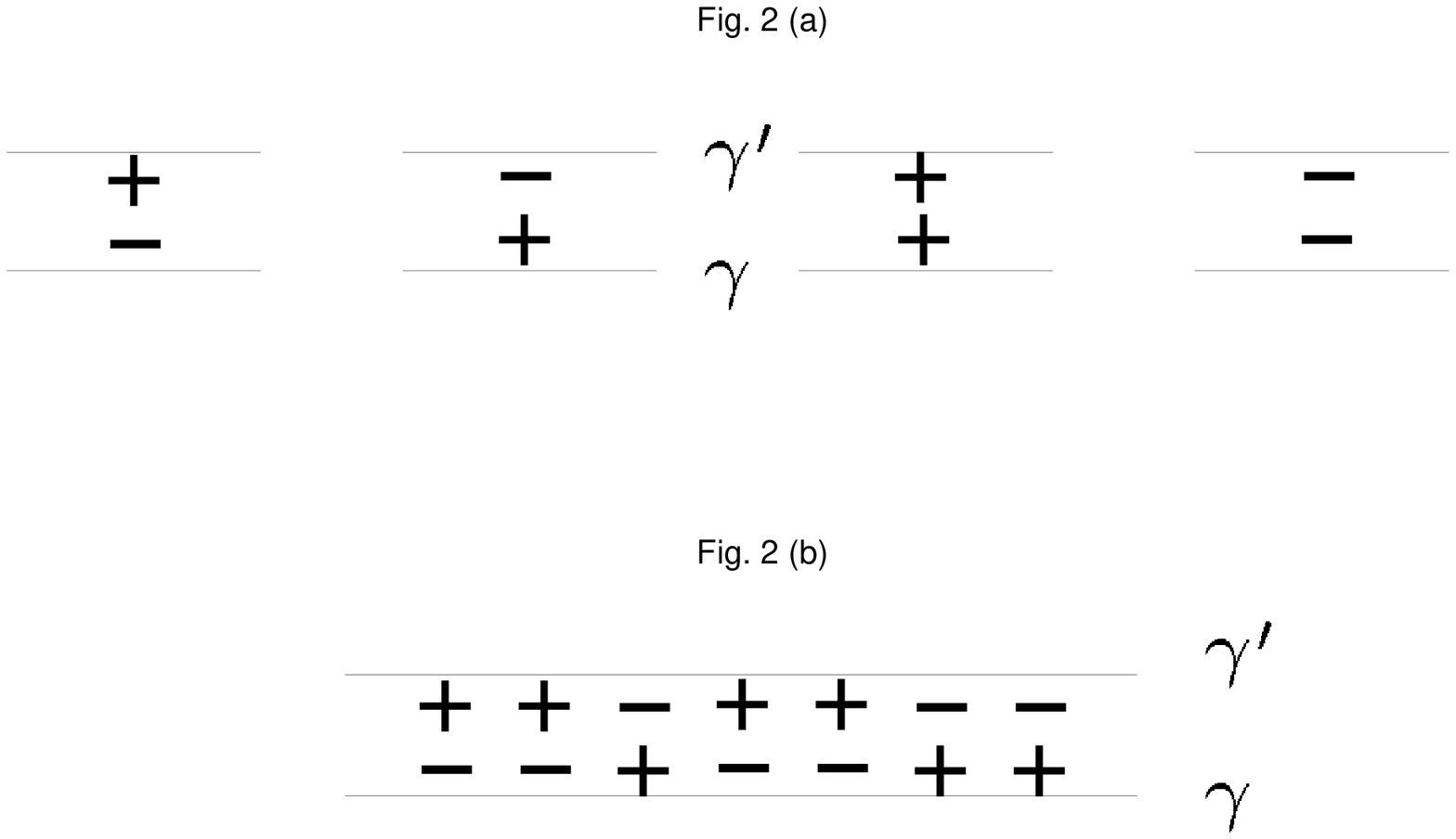}
\end{center}
\caption{(a) Allowed and forbidden encounters of orbits within a $k$-family: A self-intersection is represented by a $+$, an avoided self-crossing by $-$. Allowed encounters within a $k$-family are ${ + \choose - }$ and ${ - \choose + }$. Identical behavior of both orbits, ${ + \choose + }$ and ${- \choose -}$, does not count as one of the $k$ encounters defining the $k$-family.
(b) Example for a sequence of possible encounters in a $7$-family.}
\label{abbdelta}
\end{figure}

\vspace{.4cm}

\noindent $The \  model: \ \ $  
As has been known for a long time, geodesic flow on a surface of constant negative curvature with genus $g \geq 2$ is chaotic \cite{refR1}. According to the Gauss-Bonnet theorem, a surface of constant negative curvature $\kappa = -1$ with genus $g \geq 2$ has area $A = 4 \pi (g - 1)$. We are interested in self-intersections, because we suspect the universal behavior of classically chaotic systems to originate from the universal behavior related to the distribution of self-intersections. As has been known for a while, long closed orbits of length $v T$ (only the product of velocity $v$ and period $T$ is of relevance) have a mean number of self-intersections \cite{refM1}

\begin{equation}
\label{eqC}
N(T)  = \frac{2  v^2 T^2}{\pi A} + {\cal O} (T) \ \ .
\end{equation}

We need to work with a distribution function $P(\epsilon, T)$ for self-intersections with varying angle $\epsilon \ $  (see Fig. 3), the epsilon integral of which is the above $N(T)$, i.e. $N(T) = 2 \int_{0}^{+ \pi} P(\epsilon, T) d \epsilon$; here, a factor two arises since the distribution function is even in $\epsilon$. Sieber and Richter showed that the order-$T$ correction, not specified above, is responsible for the $\tau^2$ term in the form factor, while the order-$T^2$ term makes no contributions \cite{ref1,tobep}. They have established the distribution in question for large $T$ and in the interval $0 < \epsilon < \pi$ as

\begin{equation}
\label{eqD2}
P(\epsilon, T) = \frac{T^2 v^2}{2 \pi A} \sin   \epsilon  \  \left( 1 + \frac{4}{\lambda} \frac{\log ( \epsilon \ c)}{T}  \right) \ ; \ \ \ 
\end{equation}
for negative $\epsilon$, $P(\epsilon, T)$ is replaced by $P(- \epsilon, T)$; $c$ is a constant irrelevant for our discussion.

We now turn to the action difference for a pair of orbits arising from one intersection box. Following Richter and Sieber, we expand the action difference up to second order with respect to the intersection angle, 

\begin{equation}
\label{eqAction}
\Delta S (\epsilon, \delta_1, \delta_2) = \frac{p \epsilon}{2} ( \delta_1 + \delta_2 ) + {\cal O} (\epsilon^3) \ \ .
\end{equation}
Here, $p = m v$ denotes the absolute value of the momentum, and $\delta_1 + \delta_2$ is the closest-approach distance of the non-intersecting orbit in the close encounter; that distance is partitioned into $\delta_1$ and $\delta_2$ by the point of intersection of the self-crossing partner orbit (see Fig. 3). Near the points of closest aproach, the outer orbit turns continuously from the stable direction of the left loop of the inner periodic orbit to the stable direction of the right loop. (Incidentally, this is an intuitive reason for which the outer orbit is periodic if the inner one is). 

\begin{figure}[t]
\begin{center}
\leavevmode
\epsfxsize=0.45\textwidth
\epsffile{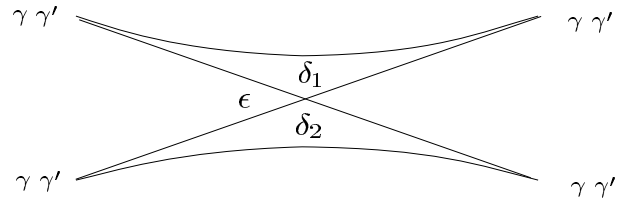}
\end{center}
\caption{Geometry of the encounter of a self-intersecting orbits with its close-by partner}  
\label{abbdelta}
\end{figure}
In the system under discussion, for large times $T_{\gamma}$, the distances approach $\delta_1 = \delta_2 \rightarrow p \epsilon / ( 2 \lambda m) $, where $\lambda$ is the Lyapunov exponent of the orbit. The Lyapunov exponent is the same for all orbits on a surface of constant negative curvature. In the limit of large $T_{\gamma}$, the action difference originating from one encounter becomes 
\begin{equation}
\Delta S ( \epsilon) = \frac{p^2 \epsilon^2}{2 m \lambda} \ \ ;
\end{equation}
that difference becomes small, eventually comparable to and even smaller than $\hbar$, for $\epsilon \rightarrow 0$.
And this is why pairs with small $\epsilon$ make non-negligible ``off-diagonal contributions'' to the form factor.  
At this point, we conclude our review of Refs.\cite{ref1,tobep}.

\vspace{.4cm}

\noindent $Multiple-loop \ contributions \ to \ the \ form \ factor: \ \ $
For the derivation of the small-$\tau$ expansion, we will have to combine the contribution of all $2^k$ different combinations of orbit pairs from the $k$-families defined in Fig. 1 to get the term $\propto \tau^{k + 1}$. 
Fig. 1 shows two different periodic orbits which are everywhere extremely close to one another except near $k$ places in configuration space where one of them self-crosses, while the other one doesn't. In analogy to the somewhat related Hikami boxes \cite{hikami} of the theory of disordered systems, we call these distinguished places in configuration space ``intersection boxes''. 
Let $\Delta S (\epsilon_j) $ be the modulus of the action difference due to $j$-th intersection box; it counts either with a positive or negative sign, to be denoted by ${\rm sgn}_j (\gamma,\gamma')$, depending on which of the two periodic orbits has the intersection. The total action difference of the two orbits is proportional to the total length difference, which stems only from the intersection boxes,

\begin{equation}
S( \gamma) - S(\gamma' ) = \sum\limits_{j=1}^{k} {\rm sgn_j}(\gamma, \gamma') \  \Delta S (\epsilon_j) \ \ .
\end{equation}

Now, consider a periodic orbit with period $T$. There are  $P(\epsilon_1, T) d \epsilon_1 $ possibilities for choosing an intersection with angle in the interval $(\epsilon_1, \epsilon_1 + d \epsilon_1)$.  We assume different intersection boxes to become uncorrelated for long periods $T$, because their number is much smaller than the total number of intersections of the periodic orbit in question. The joint distribution $P(\epsilon_1, \epsilon_2, ...., \epsilon_k, T)$ then simply involves the product of single-crossing distributions $P(\epsilon_j, T)$. Since the intersections are ordered along the orbit, we must avoid overcounting by a factor $1/k$ in 

\begin{equation}
\label{eqDF}
 P ( \epsilon_1, \epsilon_2, ..., \epsilon_k, T) = \frac{1}{k} \prod_{j = 1}^{k} P(\epsilon_j )\ \ .
\end{equation}

The contribution to the term $\tau^{k + 1}$ is due to all $2^k$ orbit pairs of all $k$-families. Let $(- \alpha, \alpha )$ be the small integration region for the intersection angles $\epsilon$, where the orbit pairs of interest exist. The semiclassical contribution of all orbit pairs from $k$ intersection boxes involves a sum over all possible $k$-families, and a sum over intra-family pairs, $K_{\rm orth}^{(k)} (\tau) \propto \sum_{\rm k-families} A_{\gamma}^2 \delta (T - T_{\gamma}) \sum_{\rm intra-family \ pairs} \exp( i S(\gamma) - i S(\gamma'))$.
The sum over intra-family pairs requires integration over the $k$ angles $\epsilon_j$ as well as summation, with the phase factor (6), over the $2^k$ possibilities of undergoing or avoiding crossings defined in Fig. 1(b),

\begin{eqnarray}
\label{eqFF}
\nonumber K_{\rm orth}^{(k)}(\tau)   = 
\lim_{\hbar \rightarrow 0}   \frac{1}{T_H}  \sum\limits_{\rm k-families} 
 A_{\gamma}^2 \delta(T - T_{\gamma})  \  \ \ \ \ \ \ \  \  \nonumber\\
\times  \int_{0}^{+ \alpha} d^k  \epsilon \left( \frac{2}{k} \prod_{j = 1}^{k} 
P ( \epsilon_j, T) \right)  \  \ \ \ \ \ \ \ 
\nonumber\\ 
\times  \sum\limits_{ \rm sign \ configurations} 
\exp \left[ \frac{i}{\hbar} \sum\limits_{l=1}^{k} {\rm sgn_l}(\gamma, \gamma') \  \Delta S(\epsilon_l) \right]   \  \ .  
\end{eqnarray}
The first sum can be evaluated using the sum rule of Hannay and Ozorio de Almeida \cite{ref0},  $\sum_{\gamma} A_{\gamma}^2  \delta(T - T_{\gamma}) \simeq T$. 

Now, observe that in  each factor ${\rm lim}_{\hbar \rightarrow 0}\int_{0}^{\alpha} d \epsilon_j P(\epsilon_j, T)$ $ \exp( \pm  i p^2 \epsilon_j^2/(2 m \lambda \hbar) ) \equiv U \pm i N$, there are two terms: A ``non-universal'' contribution $N$  which is sensitive to the sign change in the exponent, and a ``universal'' contribution $U$ insensitive to that sign change. Since the orbit pairs exist only for small $\epsilon$, we have $\sin \epsilon \simeq \epsilon$. After rescaling the integration variable as ${\tilde\epsilon} = \epsilon p/(\sqrt{2 m \lambda \hbar})$, and invoking the semiclassical limit of large $\hbar^{-1}$, the integration range can be extended to $(0, \infty)$, whereupon we find

\begin{eqnarray}
U + i N & \propto & 
4 \int_{0}^{\infty} \  d {\tilde\epsilon} \  {\tilde\epsilon} \ \left( \log {\tilde\epsilon} + ... \right) e^{\pm i {\tilde\epsilon}^2} 
\nonumber\\ 
& = & \pm i \int_{0}^{\infty} d \rho \left( \log [\rho e^{\pm i \pi/2 }] + ... \right)  \ e^{- \rho}  \nonumber\\ 
& = & - \pi/2 \pm i \ \left( \int_{0}^{\infty} d \rho \log [\rho ] \ e^{- \rho} + ... \right) \ \ .
\end{eqnarray}
The universal part $U$ comes from the term $- \pi /2$. The non-universal contribution has two components, one arising from the logarithmic correction term in $P(\epsilon, T)$ and another one, only represented by dots above, from the leading-order term $\propto T^2$ in $P(\epsilon, T)$. We thus get 

\begin{equation}
K_{\rm orth}^{k}(\tau) \propto \sum\limits_{l=0}^{k} {k \choose l} (U + i N)^l (U - i N)^{k - l} = (2 U)^k \ \ .
\end{equation}
Note that the non-universal part $N$ (which, incidentally, depends on the Lyapunov exponent and would diverge semiclassically) cancels. Sticking in all factors, we finally obtain

\begin{eqnarray}
\label{eqFF2}
 K_{\rm orth}^{k}(\tau) \  =  \ \ \ \ \ \  \ \ \ \ \ \ \ \ \ \ \ \ \ \ \ \nonumber\\ 
{\rm lim}_{\hbar \rightarrow 0} \frac{T}{T_H} \frac{2^k}{k} \left( - \frac{\pi}{2} \right)^k \left( \frac{\hbar \lambda}{m v^2} \right)^k
\left( \frac{2 v^2 T^2}{\pi A} \frac{1}{\lambda T} \right)^k \nonumber\\ 
= (-)^k \frac{2^k \tau^{k+1}}{k} \nonumber \ .\ \ \ \ \  \  \ \ \ \ \ \ \ \ \ \ \ \ \ 
\end{eqnarray}
It is reassuring to see a non-universal property like the Lyapunov exponent $\lambda$ cancel here.
By summing over $k$ we get

$$K_{\rm orth}(\tau) = 2 \tau + \tau \sum\limits_{k=1}^{\infty} \frac{(- 2 \tau )^k}{k} = 2 \tau -  \tau \log ( 1 + 2 \tau ).$$
Of course, the foregoing semiclassical series sums up to the logarithm only in the range below half the Heisenberg time, i.e. for $\tau < 1/2$, where it gives the random-matrix result; for $\tau > 1/2$, the form factor is in principle determined by its behavior up to $\tau = 1/2$ through unitarity \cite{ref0}.

By combining the present analysis with our previous work \cite{me}, the symplectic form factor is derived as

$$K_{\rm sympl.} (\tau) = \frac{\tau}{2} + \frac{\tau}{4} \sum_{k=1}^{\infty} \frac{\tau^k}{k} = \frac{\tau}{2}  - \frac{\tau}{4} \log ( 1 - \tau ).$$
Again, the series holds only up to half the Heisenberg time, which in the symplectic case means $\tau < 1$. Unitarity would have to be invoked to establish $\log(1 - \tau) \rightarrow \log | 1 - | \tau | |$, which random matrix theory claims for $| \tau| < 2$.

\vspace{.4cm}
\noindent {\it Conclusions:}
Obviously, the breakthrough achieved by Sieber and Richter has paved the way to a semiclassical understanding of the BGS-conjecture. But many open problems remain: From a physical point of view, the most intriguing one is the generalization beyond the geodesic flows considered here, in particular to ones with mixed phase spaces and to maps. Generalized time-reversal makes for another. Mathematicians will want to question much of the above, especially the assumed independence of intersections.

Difficult as the open questions just mentioned may appear, the very character of the above arguments suggests generalizability. First, we have already underscored the cancellation of the Lyapunov exponent from the universal terms. No change is therefore expected for dynamics requiring a joint distribution $P(\epsilon, \lambda, T)$ and summation over $\lambda$. Second, the existence of neighbouring pairs of unstable orbits for another type of billiards (hyperbolic boundary, flat space) was already confirmed by Sieber. The existence of a satellite to a self-intersecting orbit with small angle $\epsilon$ follows from linerization of the flow around the latter unstable orbit and should hold even when stable orbits exist elsewhere. Third, maps in two dimensional phase spaces can be embedded in four dimensional phase spaces as autonomous flows, at least if arising from Hamiltonians continuously and periodically modulated in time; projection on a two-dimensional sub-manifold should reveal self-crossing orbits and their satellites. Even the more familiar maps with Floquet operators $F = \exp ( - i H_0 / \hbar) \exp ( - i H_1 / \hbar )$ should share the relevant properties since they can be thought of as generated by Hamiltonians switching periodically between $H_0$ and $H_1$; at the instants of switching, orbits in the extended phase space will  still be continuous, but non-differentiable. The latter complication should not affect self-crossings in projections and the existence of non-self-crossing satellites.

Support by the Sonderforschungsbereich "Unordnung und gro{\ss}e Fluktuationen" of the Deutsche Forschungsgemeinschaft is gratefully acknowledged. I enjoyed helpful discussions with Martin Sieber, Uwe Abresch, Peter Braun, Gerhard Knieper and Christopher Manderfeld. Fritz Haake has put me on the right track, made many suggestions, and helped to write this paper.


\begin{thebibliography}{9}

\bibitem{ref0} F. Haake, {\it Quantum Signatures of Chaos}, 2nd edition, Springer-Verlag, Berlin, Heidelberg, 2000



\bibitem{stoeck} H.-J. St{\"o}ckmann, {\it Quantum Chaos, An Introduction}, Cambridge University Press (1999)


\bibitem{refA}
O. Bohigas,  M. J. Giannoni, and C. Schmidt, Phys. Rev. Lett. {\bf 52}, 1 (1984)


\bibitem{ref1} 
M. Sieber, K. Richter, "Correlation between Periodic Orbits and their Role in Spectral Statistics", Physica Scripta T90, 128 (2001)


\bibitem{refB}
M. V. Berry, Proc. R. Soc. London {\bf A. 400}, 229 (1985)

\bibitem{abresch}
U. Abresch, private communication


\bibitem{refR1}
Balasz, N. L. and Voros, A., Phys. Rep. {\bf 143}, 109 (1986); R. Aurich and F. Steiner, Physica D {\bf 32}, 451 (1988)



\bibitem{refM1}
S. P. Lalley, 
"Self-intersections of closed geodesics on a negatively curved surface: statistical regularities"
Convergence in ergodic theory and probability (Columbus, OH, 1993)


\bibitem{tobep}
M. Sieber, Lecture given at the Genter-Symposium, EinGedi (2001), to be published.





\bibitem{hikami}
S. Hikami, Phys. Rev. {\bf B. 24}, 2671 (1981)
\bibitem{me}
S. Heusler, "The origin of the logarithmic singularity in the symplectic form factor", nlin-0103007,
submitted to J. Phys. A.

\end{thebibliography}
\end{document}